\documentclass[fleqn,usenatbib]{mnras}

\usepackage{listings}
\usepackage{xcolor}

\usepackage{newtxtext,newtxmath}

\usepackage[T1]{fontenc}

\DeclareRobustCommand{\VAN}[3]{#2}
\let\VANthebibliography\thebibliography
\def\thebibliography{\DeclareRobustCommand{\VAN}[3]{##3}\VANthebibliography}

\usepackage{graphicx}	
\usepackage{amsmath}	

\title[Mass loss and asteroseismology]{
The Role of Asteroseismology in Understanding Mass Loss in Red Giants
}

\author[S. Örtel \& M. Yıldız]{
Sibel Örtel$^{1,2}$\thanks{E-mail: sibel.ortel@gmail.com} \thanks{ORCID: 0000-0001-5759-7790},
Mutlu Yıldız$^{1}$\thanks{ORCID: 0000-0002-7772-7641}
\\
$^{1}$Department of Astronomy and Space Sciences, Faculty of Science, Ege University, 35100, \.Izmir, Turkey\\
$^{2}$Department of Astronomy and Space Sciences, Graduate School of Natural and Applied Sciences, Ege University, 35100, \.Izmir, Turkey\\
}

\date{Accepted XXX. Received YYY; in original form ZZZ}

\pubyear{2015}

\begin{document}
\label{firstpage}
\pagerange{\pageref{firstpage}--\pageref{lastpage}}
\maketitle

\begin{abstract}
Red giant stars play a key role in advancing our understanding of stellar mass loss. However, its initial mass and the amount of mass lost during this phase remain uncertain. In this study, we investigate the asteroseismic signatures of mass loss and the parameters that influence it. We examine six stars identified as red giant branch (RGB) stars in the APOKASC-2 catalog. 
Assuming these stars are on their first ascent of the RGB, we construct interior models. The resulting model ages are significantly older than the age of the Galaxy, indicating that these stars are likely experiencing mass loss and evolving toward the red clump (RC) phase. The minimum possible initial masses are estimated using the mass–metallicity diagram, from which we infer that the minimum mass lost by these stars ranges from $0.1$-$0.3{\rm M}_{\odot}$. Models constructed with an initial minimum mass yield the maximum possible age of the star. The ages of these models fall within the range of 9–9.5Gyr. For two stars, asteroseismic parameters confirm RC classification. Due to degeneracies in the HR diagram, distinguishing between internal structure models is challenging; however, asteroseismic constraints provide clear differentiation. 
Although mass-loss and mass-conservation models have similar $M$, $R$, and $T_{\rm eff}$ values, $\Delta\nu$s determined from the $l$=0 modes in the mass-loss models are 5–10$\%$ higher than observed. This discrepancy may arise from differences in internal structure. Finally, evolutionary model grids are used to examine how initial mass and $Z$ affect mass loss. Mass loss increases with increasing metallicity and decreases with increasing initial mass, regardless of the adopted value of $\eta$.
\end{abstract}




\section{Introduction}
Asteroseismology of solar-like oscillators is one of the most effective methods for understanding the internal structure and evolution of stars. The solar-like oscillations observed on the surfaces of these stars are pressure (p-) mode waves with acoustic properties. 
These oscillation frequencies allow the determination of a star’s fundamental parameters. Moreover, since these waves carry information from various layers to the surface, they provide insight into stellar interiors. However, the amplitudes of solar-like oscillations are much smaller than those of other pulsating stars (e.g., Cepheids, Miras), making them difficult to detect with ground-based telescopes. Therefore, space-based telescopes such as the COnvection, ROtation and planetary Transits mission \citep[CoRoT;][]{Baglin2006}, \textit{Kepler} \citep{Borucki2010}, the Transiting Exoplanet Survey Satellite \citep[TESS;][]{Sullivan2015}, and the upcoming PLAnetary Transits and Oscillations of stars mission \citep[PLATO;][]{2014ExA....38..249R}, planned for launch in 2026, are crucial for detecting such oscillations. Among the key unresolved issues in stellar evolution are the mechanisms through which stars gain and lose mass. In this study, we investigate the mass-loss process in solar-like oscillating stars through the lens of asteroseismology.

{The mass ($M$) and radius ($R$) of single stars, which are generally difficult to determine, can be estimated with high precision using asteroseismic parameters derived from solar-like oscillation frequencies \citep{2007ApJ...659..616C, 2010A&A...522A...1K, 2010ApJ...713L.164C, 2015MNRAS.452.2127S, 2017ApJ...844..102H, 2022ApJ...927..167L, Yıldız2023}.}
One of these asteroseismic parameters is the large frequency separation ($\Delta\nu$), which is proportional to the square root of the mean stellar density ($\rho$) \citep{Ulrich1986}. Another key parameter is the frequency of maximum oscillation amplitude ($\nu_{\rm max}$), which scales with the surface gravity ($g$) divided by the square root of the effective temperature ($T_{\rm eff}$) \citep{Brown1991}.
Using these relations, conventional asteroseismic scaling relations \citep{Kjebed1995} are derived, from which the asteroseismic mass ($M_{\rm sca}$) and radius ($R_{\rm sca}$) can be computed as follows:
\begin{equation}
\frac{M_{\rm sca}}{M_{\sun}}=\left( \frac{\nu_{\rm max}}{\nu_{\rm max\sun}}\right)^{3} 
\left( \frac{\langle\Delta\nu\rangle}{\langle{\Delta\nu_{\sun}}\rangle}\right)^{-4} 
\left( \frac{T_{\rm eff}}{T_{\rm eff \sun}}\right)^{1.5},
\label{eq:Msca}
\end{equation}
\begin{equation}
\frac{R_{\rm sca}}{R_{\sun}}=\left( \frac{\nu_{\rm max}}{\nu_{\rm max\sun}}\right)^{} 
\left( \frac{\langle\Delta\nu\rangle}{\langle{\Delta\nu_{\sun}}\rangle}\right)^{-2} 
\left( \frac{T_{\rm eff}}{T_{\rm eff \sun}}\right)^{0.5}.
\label{eq:Rsca}
\end{equation}
In these equations, $\Delta\nu_{\sun}$, $\nu_{\rm max\sun}$, and $T_{\rm eff\sun}$ represent the solar reference values, which are $135.1\pm0.1$ $\mu{\rm Hz}$, $3090\pm30$ $\mu{\rm Hz}$, and $5772\pm0.8$ K, respectively \citep{Huber2011, Prsa2016}.
In addition to Equations \ref{eq:Msca} and \ref{eq:Rsca}, several modified scaling relations have been developed to improve the accuracy of these estimations \citep{2011ApJ...743..161W, 2011A&A...530A.142B, 2016ApJ...822...15S, 2016MNRAS.462.1577Y, 2017ApJ...843...11V, YildizOrtel2021}.

Solar-like oscillations are observed at various stages of stellar evolution, one of which is the red giant (RG) phase. The RG region of the Hertzsprung–Russell (HR) diagram includes a variety of stars: those ascending and descending the red giant branch (RGB), as well as red clump (RC) stars. RC stars are characterized by core helium burning. These evolutionary stages cannot be distinguished using non-asteroseismic properties alone.
However, asteroseismic parameters—specifically the period spacing of dipole modes ($\Delta\Pi_1$) and the large frequency separation ($\Delta\nu$)—allow us to differentiate between first-ascent RG stars and RC stars \citep{Bedding2011, Mosser2014}.
If a star has $\Delta\Pi_1 > 100$ seconds and $\Delta\nu < 10$ $\mu$Hz, it is classified as an RC star that has undergone mass loss and reached the helium-burning phase. Conversely, if $\Delta\nu < 10$ $\mu$Hz and $\Delta\Pi_1 < 100$ seconds, the star is likely on its first ascent along the RGB.
However, $\Delta\Pi_1$ is not always measurable for every star. Therefore, additional asteroseismic indicators are required to determine whether a star is losing mass during its evolution along the RGB.
{\cite{2012A&A...541A..51K} developed a method to distinguish evolved stars for which $\Delta\Pi_1$ cannot be determined. This method uses the central $\Delta\nu$ ($\Delta\nu_{\rm c}$) and the central phase shift ($\epsilon_{\rm c}$). By examining $\Delta\nu_{\rm c}$ and $\epsilon_{\rm c}$ calculated from the three central radial mode frequencies, it is possible to distinguish RGB, RC, secondary clump, and asymptotic giant branch stars from each other. Thus, for stars where $\Delta\Pi_1$ cannot be measured, the evolutionary stage can still be determined using $\Delta\nu_{\rm c}$ and $\epsilon_{\rm c}$.}

Stars on the RGB lose mass through cold stellar winds as they ascend the branch. According to the mass loss rate ($\dot{M}$) equation proposed by \cite{1975MSRSL...8..369R}, the mass loss depends on stellar radius ($R$), luminosity ($L$), mass ($M$), and a mass loss efficiency parameter ($\eta$). The equation is given as:
\begin{equation}
\label{eq:massloss}
    \Dot{M} = 4{\times}10^{-13}{\eta}L\frac{R}{M}.
\end{equation}
As a star ascends the RGB, both $R$ and $L$ increase, leading to a higher mass loss rate. A single star begins to lose mass effectively after passing the luminosity bump phase. Therefore, mass-conservation models are appropriate for stars below the luminosity bump, but not for those above it \citep{2018A&A...609A..58V}.

There are numerous studies on mass loss in cluster member stars \citep{2012MNRAS.419.2077M, 2017MNRAS.472..979H, 2020MNRAS.498.5745T, 2022MNRAS.515.3184H}. Research on RGB and RC stars in clusters investigates how mass loss occurs and how much mass is lost. Studies of field stars and open clusters indicate that stars lose approximately 0.1 $\rm M_{\sun}$ during the RGB phase \citep{2012MNRAS.419.2077M, 2021A&A...645A..85M, 2021MNRAS.501.5135Y}, whereas studies of globular clusters suggest a loss of around 0.2 $\rm M_{\sun}$ \citep{2011A&A...529A.137L, 2016A&A...590A..64S}.

\cite{2015MNRAS.448..502M} investigated the mass loss of RGB stars in globular clusters and examined the dependence of $\eta$ on metallicity. They determined $\eta$ values for each cluster and found only a weak dependence between cluster metallicity and $\eta$. Similarly, \cite{2024arXiv241008330B} showed that mass loss cannot be accurately described by a single $\eta$ value in field stars or globular clusters. Based on observational data, they found that mass loss increases with decreasing metallicity in the range [Fe/H] = –0.9 to +0.0. However, more data are needed to determine whether this trend is driven by metallicity, initial mass, or both.

A star loses part of its mass during the RGB phase before entering the RC phase. \cite{2022NatAs...6..673L} used asteroseismic methods to investigate core He-burning stars. They studied 32 stars in the RG phase with masses as low as 0.5 $\rm M_{\sun}$. The inferred mass loss for these low-mass stars exceeds 0.2 $\rm M_{\sun}$. Since a single star is unlikely to lose this much mass, it is suggested that these stars likely reached their current masses through mass transfer in binary systems.

Asteroseismic studies of RG stars are essential for understanding how structural changes occur throughout stellar evolution. This study investigates the asteroseismic signatures associated with mass loss in RG stars. For this purpose, interior models describing the mass loss processes in RG stars are constructed. The observational data used in this study are presented in Section 2. The methodology of interior modeling and the application of the {\small MESA} evolutionary code are described in Section 3. Parameters influencing mass loss in RG stars, along with the construction of the MZ diagram, are detailed in Section 4. In Section 5, the results of the models and related discussions are provided, including individual analyses of the six selected stars. Finally, the conclusions are summarized in Section 6.

\section{Asteroseismic and Spectroscopic Data for Six Red Giants}
\begin{figure}    
    \includegraphics[width=1.15\linewidth]{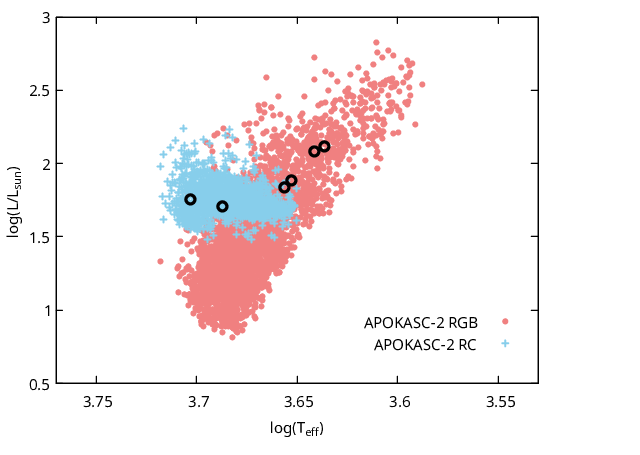}
    \caption{$\log(L/L_{\sun})$ is plotted against $\log(T_{\rm eff})$. The graph shows RGBs (solid circles) and RCs (crosses) from the APOKASC-2 catalog; the six stars studied are represented by circles.}
    \label{fig:LTeff}
\end{figure}

Spectroscopic and asteroseismic parameters of six RG stars selected from the APOKASC-2 catalogue \citep{2018ApJS..239...32P} are listed in Table \ref{tab:asteroseismic_obs}.
According to the catalogue, these stars are in the RGB phase.
Fig. \ref{fig:LTeff} shows the $\log(L/L_{\sun})$–$\log(T_{\rm eff})$ diagram for the six stars. The plot displays the positions of the selected RG stars (circles), along with RGB stars (solid circles) and RC stars (crosshairs) from the APOKASC-2 catalogue. Two of the stars are located in the region where the RC and RGB stars overlap. Two stars lie on the edge of the RC region, and the remaining two lie above it.

Spectral data for these stars are taken from APOGEE DR17 \citep{2022ApJS..259...35A}. The effective temperatures ($T_{\rm eff}$) are listed in the second column of Table \ref{tab:asteroseismic_obs}. These stars are cool red giants, with $T_{\rm eff}$ values ranging from 4330 to 5050 K. Their surface gravity ($\log{g}$) values range from 1.7 to 3.0 dex. They are metal-poor compared to the Sun, with [M/H] values between -0.77 and -0.13 dex. The fifth column of the table provides the surface heavy element abundance by mass fraction ($Z_{\rm s}$), which we calculate using the relation: $Z_{\rm s} = 10^{[\rm Fe/H]} Z_{\sun}$. Here, $Z_{\sun}$ is taken as 0.0134 \citep{2009ARA&A..47..481A}. Table \ref{tab:asteroseismic_obs} also includes $\Delta\nu$ and $\nu_{\rm max}$ values from both \cite{2019arXiv190609428K} and the APOKASC-2 catalog. The $\Delta\Pi_1$ values for two stars are available in the literature: KIC 5526130 has $\Delta\Pi_1 = 245.4$ s \citep{2016A&A...588A..87V}, and KIC 6131884 has $\Delta\Pi_1 = 255.4$ s \citep{2020A&A...639A..63G}. Observed frequencies for these six stars are taken from \cite{2019arXiv190609428K}.
\begin{table*}
    \centering
    \footnotesize 
    \setlength{\tabcolsep}{3pt} 
    \caption{Observed asteroseismic properties of six red giants. The KIC ID, $T_{\rm eff}$, $\log{g}$, [M/H], $Z_{\rm s}$, $\Delta\nu$, and $\nu_{\rm max}$ values are taken from the APOKASC-2 catalog \citep{2018ApJS..239...32P} and from \citet{2019arXiv190609428K}. The last column provides the $\Delta\Pi_1$ values compiled from the literature. The $\Delta\Pi_1$ values of the first two rows of stars are taken from \citet{2016A&A...588A..87V} and \citet{2020A&A...639A..63G}.}
    \begin{tabular}{clccccrcrc}
        \hline   
         & &&&&\multicolumn{2}{c|}{APOKASC-2} & \multicolumn{2}{c|}{\cite{2019arXiv190609428K}} \\
        KIC& {$T_{\rm eff}$} & {$\log{g}$} & {M/H} & $Z_{\rm s}$& $\langle\Delta\nu\rangle$ & {${\nu_{\rm max}}$ }& $\langle\Delta\nu\rangle$ & {${\nu_{\rm max}}$ }& $\Delta\Pi_{1}$ \\
         &  (K) & (cgs) &&& (${\mu \rm Hz}$) & (${\mu \rm Hz}$) & (${\mu \rm Hz}$) & (${\mu \rm Hz}$) & (s) \\
        \hline
        5526130& $5049.1\pm21.3$ & $2.330\pm0.046$ & $-0.773\pm0.013$ & 0.0023 & $3.567\pm0.011$ & $23.282\pm0.014$ & $3.492\pm0.019$ & $23.330\pm0.187$ &   $245.4\pm8.94$\\
        6131884& $4866.7\pm10.0$ & $2.970\pm0.029$ & $-0.612\pm0.007$ & 0.0033 & $3.703\pm0.010$ & $27.018\pm0.020$ & $3.612\pm0.010$ & $26.858\pm0.282$ & $255.4$\\
        6521069& $4330.9\pm8.8$ & $1.767\pm0.030$ & $-0.297\pm0.010$ & 0.0068 & $1.289\pm0.011$ & $7.070\pm0.067$ & $1.294\pm0.074$ & $7.178\pm0.090$ & --- \\
        7188156& $4496.9\pm11.9$ & $1.928\pm0.039$ & $-0.705\pm0.012$ & 0.0026 & $2.122\pm0.007$ & $13.547\pm0.027$ & $ 2.125\pm0.004$ & $13.888\pm0.127$ & ---\\
        7668536& $4533.4\pm6.9$ & $2.094\pm0.023$ & $-0.136\pm0.006$ & 0.0098 & $2.356\pm0.018$ & $15.149\pm0.043$ & $2.376\pm0.002$ & $15.157\pm0.214$ & ---\\
        8081853& $4382.4\pm6.3$ & $1.842\pm0.022$ & $-0.143\pm0.007$ & 0.0096 & $1.332\pm0.007$ & $7.064\pm0.045$ & $1.349\pm0.007$ & $6.768\pm0.136$ & ---\\
        \hline
    \end{tabular}
    \label{tab:asteroseismic_obs}
\end{table*}

\section{Modelling Method}
The Modules for Experiments in Stellar Astrophysics \citep[{\small MESA};][]{Paxton2011, Paxton2013, Paxton2015, Paxton2018, Paxton2019, Jermyn2023} evolution code is used to construct interior models. The basic input parameters required to construct a model are $M$, initial metallicity ($Z_0$), initial helium abundance ($Y_0$), and the mixing length parameter ($\alpha$). 

%
%
{The most important fundamental parameter determining the evolutionary process of a star is its $M$. We calculate $M$ and $R$ of stars exhibiting solar-like oscillations using the scaling relations. Another key parameter affecting stellar evolution is $Z$. In evolved stars, heavy elements that settle in the deeper layers during the MS phase are brought back to the surface by convection, so the initial metallicity $Z_0$ is taken to be equal to the surface metallicity $Z_s$ obtained from [Fe/H]. The initial helium abundance is related to $Z_0$ through $Y_0 = 2Z_0 + 0.2471$, where 0.2471 represents the primordial helium abundance \citep{2020A&A...641A...6P}. Another input parameter is the mixing-length parameter ($\alpha$), which is varied to fit the model radius to the radius obtained from the scaling relations.}  

{Equation (\ref{eq:massloss}) is used in the MESA evolution code to describe mass loss in red giant branch stars. In this framework, the mass-loss rate is controlled by the efficiency parameter $\eta$.}  

{Once an interior model consistent with the observational position of a star in the HR diagram is obtained, the oscillation frequencies are calculated. The model value of $\Delta\nu$ is then compared with the observed one. If the two values disagree within the observational uncertainties, the stellar mass is recalculated from the $\Delta\nu$–$\rho$ relation while keeping $R$ fixed, and the model is reconstructed with this updated mass. This procedure is repeated iteratively until satisfactory agreement with the observed $\Delta\nu$ is achieved.}

\subsection{Properties of {\small {MESA}} inlist}
\label{sec:MESA}
{In this study, version r23.05.1 of the MESA evolution code is used to construct interior models.}
The interior models include the pre-MS phase. The Nuclear Astrophysics Compilation of Reaction Rates (NACRE) tables \citep{Angulo1999} are used for nuclear reaction rates. 
{Opacities are taken from the OPAL tables \citep{Iglesias1993, Iglesias1996} at high temperatures and from the \cite{Ferguson2005} tables at low temperatures, both of which are consistent with the solar abundances of \cite{2009ARA&A..47..481A}.}
Element diffusion is included in the models; for this, the method given by \cite{1986ApJS...61..177P} is used. The \texttt{photosphere} option is used in interior models for atmospheric conditions that affect the adiabatic oscillation frequencies. 
{For mass loss on the RGB, we adopt the \citet{1975MSRSL...8..369R} mass-loss relation.} 
{An example \texttt{inlist} used for model construction is given in the appendix.}

The adiabatic oscillation frequencies of the interior model are calculated with the red-giant package in the {\small ADIPLS} \citep{Christensen2008}. For surface-effect correction, the method of \cite{KBC2008} is applied to the model frequencies.

\section{Red giant stars}
\label{RGs}
Low mass stars cannot ignite helium directly after they exhaust the hydrogen in their core. Once the hydrogen in the core is exhausted, hydrostatic equilibrium is disrupted, {and the star contracts}. Just above the core, the hydrogen shell ignites when the temperature reaches values that can burn hydrogen. This hydrogen shell burning creates a mirror effect: a star's envelope expands as its core collapses.

As a star climbs the RGB, the convective layer gradually descends deeper.
When the convective layer exceeds the main sequence core boundary, the mirror effect disappears, {the star undergoes contraction}, and its luminosity and effective temperature decrease \citep{2020MNRAS.492.5940H}. 
When the mean molecular weight discontinuity is reached in the hydrogen-burning shell, the mirror effect becomes active again, and the star continues its climb \citep{2015MNRAS.453..666C, 2020MNRAS.492.5940H}. This phase is called the luminosity bump in the HR diagram and corresponds to the region in the range $\log(L/{\rm L_{\sun}})=1.5-2.0$.

Once a star passes its luminosity bump, it effectively begins to lose mass via cold stellar winds. However, the most significant mass loss occurs around the RGB tip. The mass loss depends on $M$, $L$, $R$ and $\eta$ in the Reimers' mass loss rate \citep{1975MSRSL...8..369R} (equation \ref{eq:massloss}). According to our investigations, the $Z_0$ value needs to be included among these fundamental parameters (see Section \ref{sec:M0Zs}). 

A star continues to rise, losing mass in the RGB until a helium flash occurs in its degenerate core. 
After the helium flash, the core expands. The increase in core radius pushes the hydrogen shell upwards, causing the temperature to decrease. This causes the energy released as a result of the burning of the hydrogen shell to decrease. Therefore, there is not enough energy transfer to the upper layers, and the envelope contracts rapidly, moving towards RC on the HR diagram. 
The star then undergoes a series of helium flashes in the RC, after which helium burning begins in the core.

\subsection{Mini Helium Flashes}
\label{sec:mHeF}
When a star ascends to the tip of the RGB, a main helium flash is ignited in a specific shell of the degenerate core. Following this event, the star rapidly evolves toward the RC region. However, upon reaching the RC region, it does not immediately settle into stable core helium burning, as the core remains largely degenerate. To lift this degeneracy, the star undergoes a series of so-called "mini helium flashes". 

\begin{figure*}
    \centering
    \includegraphics[width=1.0\linewidth]{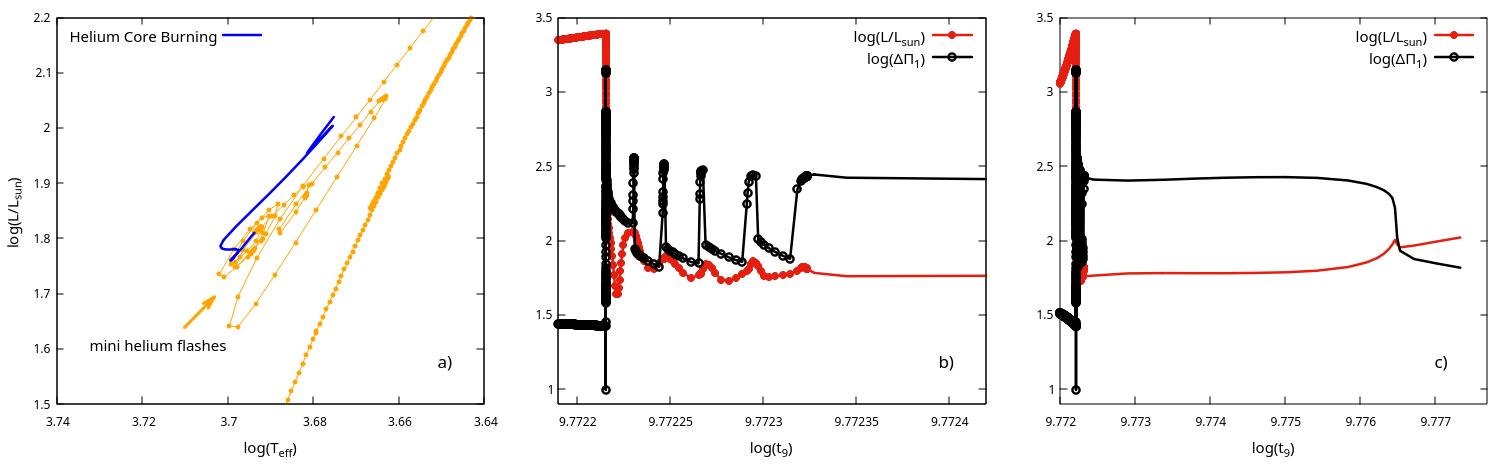}
    \caption{$L$ and $\Delta\Pi_1$ variations of an internal structure model constructed for {KIC 5526130}. (a) The HR diagram shows the track followed by the star when it reaches the RC region. (b) $L$ and $\Delta\Pi_1$ are plotted logarithmically against age. Here, the change during the mini helium flash is shown. (c) The $L$ and $\Delta\Pi_1$ changes in the phase where regular helium burning starts after the mini helium flash are given.}
    \label{fig:RC_triple}
\end{figure*}
\begin{figure}
    \centering
    \includegraphics[width=1.05\linewidth]{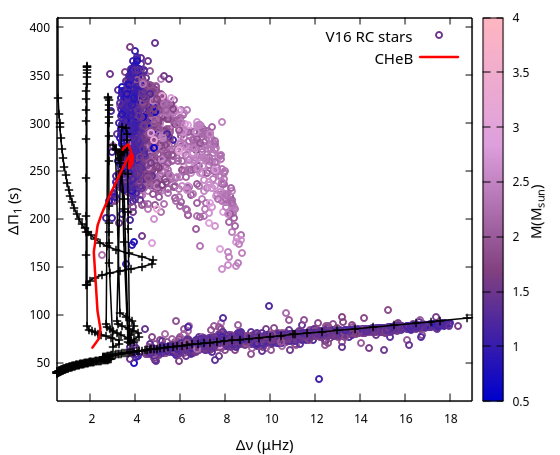}
    \caption{$\Delta\Pi_1$–$\Delta\nu$ diagram. The 1.05 $M_\odot$ model shows the evolutionary path of the stellar interior. The pre-ZACHeB phase is marked with black crosses, while the stable core helium-burning phase is shown as a red solid line. The background circles represent the values determined by Vrard et al. (2016) for 6111 stars, with the color variation indicating stellar mass.}
    \label{fig:RC_DP1-Dnu}
\end{figure}

These flashes are episodic in nature. On the HR diagram, this transitional phase manifests as a sequence of nested loops. Fig. \ref{fig:RC_triple}a shows the evolutionary tracks during the mini helium flashes and the stable core helium burning.
Fig. \ref{fig:RC_triple}b displays the temporal evolution of $L$ and $\Delta\Pi_1$ during a mini helium flash. Both $L$ and $\Delta\Pi_1$ exhibit a gradual rise followed by a sharp decline in tandem during each mini helium flash, indicating their close link to the core’s structural changes induced by the flash events. During these flashes, $\Delta\Pi_1$ can rise to $\sim350$ s, subsequently declining to $\sim70$ s after the flash concludes. Fig. \ref{fig:RC_triple}c shows the stable helium burning phase. At Zero Age Core Helium Burning (ZACHeB), the $\Delta\Pi_1$ value is approximately 280 s. It then rapidly drops to about 60 s as the star approaches Terminal Age Core Helium Burning (TACHeB) (also see Fig. \ref{fig:RC_DP1-Dnu}).

A star in the mini helium flash phase may be misclassified as either an RC star or a first-ascent RGB star, depending on the timing of the observation. Fig. \ref{fig:RC_DP1-Dnu} shows the evolution of a model on the $\Delta\nu$–$\Delta\Pi_1$ diagram. If a star is observed during the interval between two mini helium flashes, it may be misclassified as a first-ascent RGB star. Conversely, if it is observed during a mini helium flash, it may be classified as an RC star. While classification as an RC star is not inaccurate, given the star’s rapid evolution toward stable core helium burning, identifying it as a first-ascent RGB star complicates the construction of an accurate interior model. However, this phase is extremely short compared to both the stable helium-burning phase and the first-ascent RGB phase. As a result, the number of stars observed in this stage is expected to be only a few among thousands of RG stars.

\subsection{Effects of Metallicity and Initial Mass on Stellar Mass Loss} 
\label{sec:M0Zs}
In RGB, the mass loss rate depends on the fundamental stellar parameters. To determine which parameters this change depends on, we construct evolution grids using the variables $M_0$, $Z_0$ and $\eta$. The models are constructed with the evolution code {\small MESA}, and the {\tt cool\_wind\_RGB\_scheme = 'Reimers'} option is used for mass loss. $M_0$ varies between 1.0 to 1.4 in 0.1 steps and $Z_0$ varies between 0.005 to 0.035 in 0.005 steps. Grids are constructed using three different mass loss parameters, $\eta=0.20$, 0.50 and 0.85. 
The value of $Y_0$ is computed from the relation $Y_0=2Z_0+0.2471$. Finally, $\alpha$ is taken as the solar value ($\alpha_{\sun}=1.8311$).

\begin{figure}
	\includegraphics[width=\columnwidth]{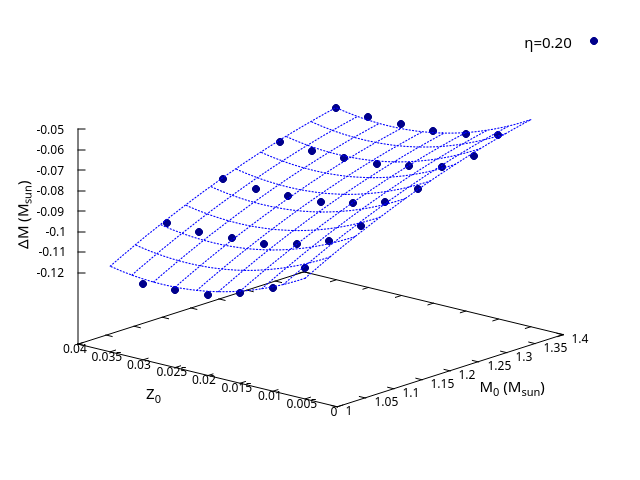}
    \includegraphics[width=\columnwidth] {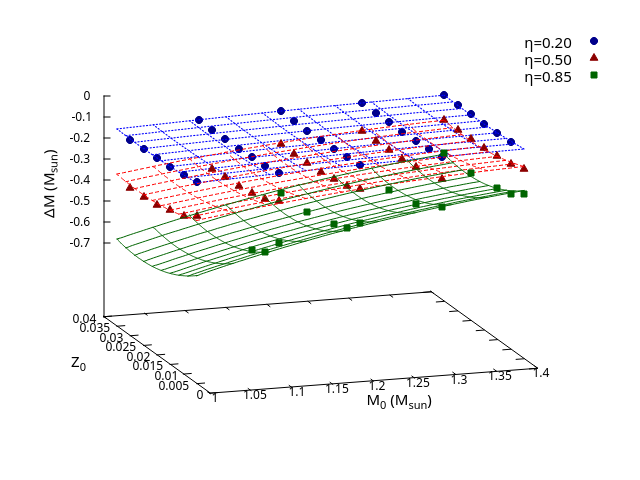}
    \caption{Initial mass ($M_0$) and metallicity ($Z_0$) are plotted against the amount of mass loss ($\Delta M$). The top panel shows the models constructed with $\eta=0.20$ (filled circle). As $M_0$ decreases, the mass loss increases, while as $Z_0$ increases, the mass loss increases. In the bottom panel, models constructed with $\eta=0.20$ (filled circle), 0.50 (filled triangle) and 0.85 (filled square) are given together. The effects of $M_0$, $Z_0$ and $\eta$ on mass loss are investigated together.}
    \label{fig:DMZMi}
\end{figure}
The total amount of mass loss in RGB ($\Delta M_{\rm RGB}=M-M_0$) varies depending on the values of $M_0$ and $Z_0$, regardless of the mass loss parameter $\eta$. The relationship between $M_0$, $Z_0$ and $\Delta M_{\rm RGB}$ is shown in 3D in Fig. \ref{fig:DMZMi}. In the top panel, $M_0$ and $Z_0$ of the models with $\eta=0.2$ are plotted against $\Delta M_{\rm RGB}$.
As $Z_0$ increases, the total amount of mass loss increases. When $Z_0$ is approximately 0.0300, maximum mass loss occurs, and saturation can be said to be reached.
Increasing $M_0$ reduces the total mass loss. These parameters determine how long the star will linger near the RGB tip, where it loses mass most effectively. In this case, the model that loses the most mass by spending more time in the RGB tip than the other models has the lowest $M_0$ and the highest $Z_0$ value.

In the bottom panel of Fig. \ref{fig:DMZMi}, $M_0$ and $Z_0$ of models with different $\eta$ values are plotted against $\Delta M$. Here, the effect of $\eta$ on mass loss is clearly seen. Increasing $\eta$ increases the total mass loss for each case. However, the change in mass loss due to $\eta$ is not linear. As we go to lower masses, the effect of $\eta$ on mass loss increases.
Fig. \ref{fig:DMZMi} shows the surfaces defined for each value of $\eta$. Using these defined surfaces, it is possible to calculate the initial mass of the star for each $\eta$. 
To use this relation, $M_{\rm sca}$ and $Z$ are required. For each $\eta$, $M_0$ can be calculated using the following equations:
\begin{equation}
\label{eq:etas}
    \begin{split}
M_{0,\eta=0.20}&=0.28+0.70M+0.09M^2+1.97Z_0-36.57Z_0^2,\\
\\
M_{0,\eta=0.50}&=0.64+0.29M+0.24M^2+4.89Z_0-90.02Z_0^2,\\
\\
M_{0,\eta=0.85}&=0.81+0.24M+0.22M^2+7.66Z_0-141.23Z_0^2.\\
    \end{split}
\end{equation}

\subsection{Mass-Metallicity diagram}
\begin{figure}
    \includegraphics[width=1.15\linewidth]{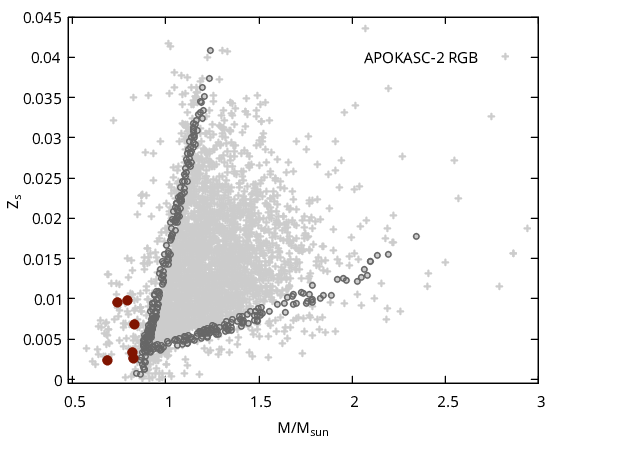}
    \caption{$Z_s$ is plotted  against $M$. The six stars examined are shown as filled circles. In the background, RGB stars from the APOKASC-2 catalog are represented by crosses, and the two sides of the triangle are indicated by open circles \citep{Yıldız2023}. 
    All six stars lie to the left of this boundary.}
    \label{fig:MZ}
\end{figure}

Mass and $Z$ are the two most important fundamental parameters that determine the course of stellar evolution. In \cite{Yıldız2023}, the mass-metallicity (MZ) diagram of RG stars in the APOKASC-2 catalog is given. There is a triangular structure in the MZ diagram (Fig. \ref{fig:MZ}).   
Stars at the base of the triangle have low $Z_{\rm s}$ and relatively high $M$. 
The evolution of a star with low $Z_{\rm s}$ and relatively high mass will be rapid. In this case, at the base of the triangle are the stars that are preparing to leave the RGB phase.
Above the left edge are the minimum mass stars that can reach the RGB phase, depending on $Z_{\rm s}$. The stars above this edge are the oldest stars in the RGB phase.
As seen in Fig. \ref{fig:MZ}, as $Z_{\rm s}$ increases, $M$ also increases. Because increasing Z slows down evolution. 
Therefore, in order to reach RGB, $M$ needs to increase. 

In Fig. \ref{fig:MZ}, low-mass stars located outside the left edge of the triangle take longer to reach the RGB. Their stay outside the triangle can be explained by mass loss. The minimum initial mass ($M_{0{\rm min}}$) of these stars can be taken as $M$ on the left edge with the same $Z$.
Table \ref{tab:models} gives the masses $M_{0{\rm min}}$ determined in this way for the six stars studied in this paper. Estimated $\eta$ values for these initial masses are obtained using the equations (\ref{eq:etas}).

\section{Model results and Discussions}
\label{sec:ResandDis}
\begin{figure*}
    \centering
    \includegraphics[width=1.0\linewidth]{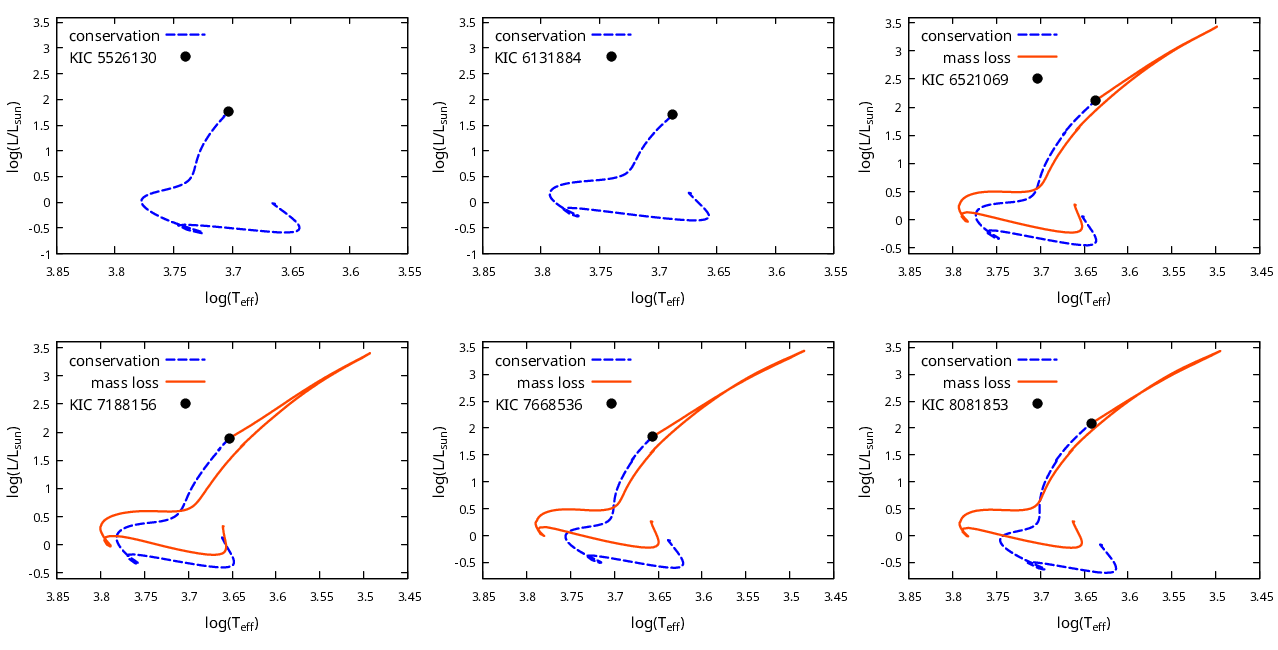}
    \caption{$\log(L)-\log(T_{\rm eff})$ plot of mass-conservative and mass-loss models. Since KIC 5526130 and KIC 6131884 are RC stars, only mass-conservative models are constructed. Mass-conservative and mass-loss models are represented by dashed lines and solid lines, respectively. The solid black circle shows the position of the star on $\log(L)-\log(T_{\rm eff})$.}
    \label{fig:sixHRD}
\end{figure*}
\begin{table*}
    \centering
    \caption{The basic properties of mass-conservative and mass-loss models. The columns show the KIC ID, model types, final mass ($M_{\rm f}$) and radius ($R_{\rm f}$) of the model, model effective temperature ($T_{\rm eff}$), minimum initial mass ($M_0{\rm min}$), minimum mass loss ($\Delta M_{\rm min}$), initial metallicity ($Z_0$) and helium abundance ($Y_0$), convective parameter ($\alpha$), mass loss parameter ($\eta$), age ($t_9$), large separation ($\Delta\nu$) and frequency of maximum amplitude ($\nu_{\rm max}$), respectively.}
    \begin{tabular}{cccccccccccccc}
        \hline   KIC&model&$M_{\rm f}$&$R_{\rm f}$&$T_{\rm eff}$&$M_{\rm 0,min}$&$\Delta M_{\rm min}$&$Z_{0}$&$Y_{0}$& ${\alpha}$&$\eta_{\rm min}$ &$t_9$ & $\langle\Delta\nu\rangle$& ${\nu_{\rm max}}$ \\[1.2pt]
        &&(${M_{\sun}}$)&(${R_{\sun}}$)&(K)&(${M_{\sun}}$)&(${M_{\sun}}$)& & & & &(Gyr)&(${\mu Hz}$)&(${\mu Hz}$) \\[1.2pt]
        \hline
        5526130& conservation & 0.6878 & 9.8686 & 5048 & 0.6878 &0& 0.0023 & 0.2517 & 2.3979 & 0 & 24.369 & 3.4430 & 23.420\\
        & mass loss & --- & --- & --- & 0.92 &0.2322& 0.0023 & 0.2517 & --- & 0.4083 & --- & --- & --- \\
        6131884& conservation & 0.8219 & 10.0826 & 4867 & 0.8219 &0& 0.0033 & 0.2537 & 2.0604 & 0 & 14.150 & 3.7159 & 27.306\\
        & mass loss & --- & --- & --- & 0.93 &0.1081& 0.0033 & 0.2537 & --- & 0.2077 & --- & --- & --- \\
        6521069& conservation & 0.8331 & 20.2957 & 4333 & 0.8331 &0& 0.0068 & 0.2603 & 1.9773 & 0 & 16.138 & 1.3251 & 7.239 \\
        & mass loss & 0.8348 & 20.2633 & 4335 & 0.97 &0.1352& 0.0068 & 0.2603 & 1.7110 & 0.249 & 9.538 & 1.3873 & 7.276 \\
        7188156& conservation & 0.7874 & 14.4948 & 4497 & 0.7874 &0& 0.0026 & 0.2523 & 1.6537 & 0 & 15.751 & 2.1754 & 13.168\\
        & mass loss & 0.7883 & 14.4896 & 4497 & 0.92 &0.1317& 0.0026 & 0.2523 & 1.3168 & 0.224 & 9.202 & 2.3532 & 13.193\\
        7668536& conservation & 0.7915 & 13.4059 & 4533 & 0.7915 &0& 0.0098 & 0.2667 & 2.2063 & 0 & 21.751 & 2.3978 & 15.413\\
        & mass loss & 0.7924 & 13.4085 & 4532 & 1.02 &0.2276& 0.0098 & 0.2667 & 1.7811 & 0.406 & 9.097 & 2.5451 & 15.426\\
        8081853& conservation & 0.7425 & 19.1355 & 4384 & 0.7425 &0& 0.0096 & 0.2663 & 2.2810 & 0 & 26.825 & 1.3578 & 7.216\\
        & mass loss & 0.7524 & 19.1128 & 4382 & 1.01 &0.2576& 0.0096 & 0.2663 & 1.9061 & 0.469 & 9.327 & 1.4289 & 7.331 \\
        \hline
    \end{tabular}
    \label{tab:models}
\end{table*}
RGB stars lose mass during their ascent due to cold stellar winds. 
{As a star ascends the RGB, the rate of mass loss increases. In the evolutionary grids, the most significant mass loss is seen near the RGB tip.}
The six stars studied in the paper are located around the luminosity bump on the HR diagram. Two different evolution scenarios can be considered for these stars. Let's assume that stars are in first ascent. After the luminosity bump, mass loss begins to take effect, and it is most effective around the RGB tip. Therefore, mass-conservative models of stars can be constructed. The other scenario is that the star ascends to the RGB tip, loses mass, and then descends towards the RC region. For this, mass-loss models are constructed. The basic properties of the models for both cases are given in Table \ref{tab:models}.

In the first scenario, the age of the models is much higher than the estimated age of our galaxy \citep[][$13.4\pm0.8$ Gyr]{2003A&A...408..529G}. This means that more time is needed to reach the RG region with these $M$s and $Z$s.  
Therefore, these six stars must lose mass to be located in the RG region. Accordingly, the models assume mass loss in the second scenario. The minimum mass loss amounts ($\Delta M_{\rm min}$), determined from the MZ diagram for the six stars, range between 0.10 and 0.26 ${\rm M}_{\sun}$.
$M_{0\rm min}$s range from 0.92 to 1.02 $\rm M_{\sun}$. $\eta$ and $\alpha$ values are between 0.20-0.50 and 1.3-2.0, respectively. The ages of the mass loss models constructed with these values give the maximum age for the stars. Model ages are around 9-9.5 Gyr. 

In mass-conservative models, the initial mass $M_0$ is taken as $M_{\rm sca}$, which is computed via the scaling relation. In mass-loss models, agreement is achieved between the final mass $M_{\rm f}$ and $M_{\rm sca}$. For KIC 6521069, KIC 7188156, and KIC 7668536, there is a 0.2$\%$ agreement between the $M_{\rm f}$ values of the models constructed using both approaches, while for KIC 8081853 the agreement is 1.4$\%$. The final radius $R_{\rm f}$ is also in very good agreement between the two types of models, with differences less than 0.2$\%$.

The $\log(L)-\log(T_{\rm eff})$ graph of all the models is given in Fig. \ref{fig:sixHRD}. Mass-loss (line) and mass-conservative (dashed line)  models are fitted to the same point on the HR diagram. The evolutionary tracks of both cases are shown for four stars. In the figure, only the first ascent phase is shown for KIC 5526130 and KIC 6131884. Although these stars are listed as RGB in the APOKASC-2 catalogue, their $\Delta\Pi_{1}$ have been detected in the literature and identified as RC stars. RC models have been constructed for these two stars (see below).

\subsection{Red Clump Models for KIC 5526130 and KIC 6131884}
\label{sec:RCmodel}
Most RC stars lie in the same region of the HR diagram as RGB stars (see Fig. \ref{fig:LTeff}). Two of the six stars we studied are located in this region. The $\Delta\Pi_1$ values of these stars have been determined in the literature. According to the determined $\Delta\Pi_1$ and $\Delta\nu$ values, KIC 5526130 and KIC 6131884 are in the RC phase (see Table \ref{tab:asteroseismic_obs}). Therefore, RC models have been constructed for these two stars.
\begin{table*}
    \centering
    \caption{Basic properties of the RC models.}
    \begin{tabular}{lccccccccccccc}
        \hline   
        Model ID & $M_{\rm f}$ & $M_{\rm 0}$ & $\Delta{M}$ & $R_{\rm f}$ & $\eta$ & $\alpha$ & $t_9$ & $\Delta\nu $ & $\nu_{\rm max}$ & $\Delta\Pi_1$& $T_{\rm eff} $ &$\chi2_{\rm seis}$& evolution  \\
         & $(\rm M_{\sun})$ & $(\rm M_{\sun})$ & $(\rm M_{\sun})$ & $(\rm R_{\sun})$ &  &  & (Gyr) & $(\mu{\rm Hz})$ & $(\mu{\rm Hz})$ & ({\rm s})& $({\rm K})$ & & stage\\
        \hline
        \multicolumn{11}{l|} {KIC 5526130} \\
        \hline
        RC1A & 0.7080 & 0.93 & 0.2220 & 10.37 & 0.45 & 1.7811 & 9.0790 & 3.554 & 21.827 & 280.4 & 5104 & 8.36 & bZACHeB \\
        RC1B & 0.8563 & 1.05 & 0.1937 & 10.98 & 0.45 & 1.6311 & 5.9200 & 3.572 & 23.916 & 271.5 & 4939 & 4.36 & ZACHeB \\
        \textbf{RC1C} & 0.8498 & 1.05 & 0.2002 & 11.05 & 0.45 & 1.6311 & 5.9692 & 3.571 & 23.38 & 250.2 & 4978 & 0.21 & \textbf{nTACHeB} \\
        RC1D & 0.8818 & 1.08 & 0.1982 & 11.09 & 0.50 & 1.8311 & 5.4142 & 3.561 & 23.840 & 238.8 & 5081 & 0.42 & nTACHeB \\
        {\cite{2019arXiv190609428K}} & --- & --- & --- & --- & --- & --- & --- & 3.492 & 23.330 & 245.4 & 5049.1 & --- & --- \\
        &  &  &  &  &  &  &  & $\pm$0.019 & $\pm$0.187 & $\pm$8.94 & $\pm$21.3 & & \\
        APOKASC-2 & --- & --- & --- & --- & --- & --- & --- & 3.567 & 23.282 & 245.4 & 5049.1 & --- & --- \\
        &  &  &  &  &  &  &  & $\pm$0.011 & $\pm$0.014 & $\pm$8.94 & $\pm$21.3 & & \\
        \hline
        \multicolumn{11}{l|} {KIC 6131884} \\
        \hline
        RC2A & 0.8511 & 1.08 & 0.2289 & 10.85 & 0.50 & 1.5552 & 5.6286 & 3.627 & 24.634 & 276.8 & 4838 & 3.95 & nZACHeB\\
        RC2B & 0.8510 & 1.08 & 0.2290 & 10.58 & 0.50 & 1.5552 & 5.6287 & 3.762 & 25.828 & 267.5 & 4859 & 18.31 & nZACHeB\\
        RC2C & 0.8448 & 1.08 & 0.2352 & 10.88 & 0.50 & 1.5552 & 5.6710 & 3.640 & 24.213 & 262.2 & 4880 & 4.21 & nTACHeB\\
        \textbf{RC2D} & 0.8446 & 1.08 & 0.2354 & 10.94 & 0.50 & 1.5552 & 5.6721 & 3.613 & 23.965 & 261.0 & 4876 & 0.20 & \textbf{nTACHeB}\\
        RC2E & 0.8445 & 1.08 & 0.2355 & 10.99 & 0.50 & 1.5552 & 5.6730 & 3.586 & 23.727 & 259.8 & 4873 & 3.50 & nTACHeB\\
        {\cite{2019arXiv190609428K}} & --- & --- & --- & --- & --- & --- & --- & 3.612 & 26.858 & $255.4$ & 4866.7 & --- & ---\\
        &  &  &  &  &  &  &  & $\pm0.010$ & $\pm0.282$ &  & $\pm$10 & & \\
        APOKASC-2 & --- & --- & --- & --- & --- & --- & --- & 3.703 & 27.018 & $255.4$ & 4866.7 & --- & --- \\
        &  &  &  &  &  &  &  & $\pm0.010$ & $\pm0.020$ &  & $\pm$10 & & \\
    \hline
    \end{tabular}
    \label{tab:RCmodels}
\end{table*}

The compatibility of the RC models with the observational values of $\Delta\nu$, $\Delta\Pi_1$, and $T_{\rm eff}$ is taken into account during their construction. To assess the agreement between the models and the observations, the $\chi^2$ method is applied to the seismic data of $\Delta\nu$ and $\Delta\Pi_1$. The seismic $\chi^2$ ($\chi^2_{\rm seis}$) of the models is calculated using the following expression:
\begin{equation}
{\chi^2_{\rm seis}}={\frac{1}{2}}{\sum\limits_{i=1}^{2}{\left({\frac{{f_{i\rm obs}}-{f_{i\rm mod}}}{e_{i \rm obs}}}\right)^2}}
\label{eq:chisquare.seis}
\end{equation}
where $f_{i\rm obs}$ are the observed values of $\Delta\nu$ and $\Delta\Pi_1$, $f_{i\rm mod}$ are the corresponding model predictions, and $e_{i\rm obs}$ are the observational uncertainties. {The calculated model frequencies are corrected for surface effects (see Section \ref{sec:MESA}).}

The mass-loss parameter $\eta$ is set to 0.45 and 0.50. The basic properties of the RC models for these two stars are given in Table \ref{tab:RCmodels}. The last column of the table indicates the evolutionary stage of the models. These stages are defined as: Zero Age Core Helium Burning (ZACHeB), before ZACHeB (bZACHeB), near ZACHeB (nZACHeB), and near Terminal Age Core Helium Burning (nTACHeB). In the bZACHeB phase, the star undergoes a series of mini helium flashes (see Section \ref{sec:mHeF}).

For KIC 5526130, three different $M_0$ values are used for the interior models. In the RC1A model, $M_0$ is 0.93 $\rm M_{\sun}$, determined from the MZ diagram, and its evolution corresponds to the last mini helium flash. The mass loss parameter $\eta$ is set to 0.45. $M_{\rm f}$ of the model is 0.7080 $M_{\odot}$, with a total mass loss of 0.2220 $M_{\odot}$. The final mass agrees well with the mass of approximately 0.69 $M_{\sun}$ predicted by $M_{\rm sca}$. The $\Delta\nu$ value of 3.554 $\mu$Hz is close to the value given in the APOKASC-2 catalogue, within the error limits. However, the $\Delta\Pi_1$ value is 280 s and the $T_{\rm eff}$ value is 5014 K, both of which are considerably higher than the observational values. Among the four models, this one has the highest chi-square value, $\chi^2_{\rm seis} = 8.36$.

In RC1B and RC1C, $M_0$ is 1.05 $M_{\odot}$ and $\eta$ is 0.45. In RC1D, $M_0$ is 1.08 $M_{\odot}$ and $\eta$ is 0.50. The $\Delta\nu$ value of all three models is in very good agreement with the APOKASC-2 value. The $\Delta\Pi_1$ value of RC1B is 271.5 s, which is above the value reported by Vrard et al. (2016). The evolutionary stage of this model is ZACHeB. The $\Delta\Pi_1$ values of RC1C and RC1D are within the observational error limits. The evolutionary stages of these models are nTACHeB. However, the $T_{\rm eff}$ values of all three models lie outside the APOKASC-2 uncertainty range. According to the SIMBAD database \citep{2000A&AS..143....9W}, the reported $T_{\rm eff}$ values for this star range from 4923 K to 5049 K. Considering $\Delta\nu$, $\Delta\Pi_1$, and $T_{\rm eff}$ together, RC1C appears to be the best-fitting model for KIC 5526130. At the same time, this model yields the lowest $\chi^2_{\rm seis}$ value, with $\chi^2_{\rm seis} = 0.21$. $M_{\rm f}$ values of the models are higher than the value predicted by $M_{\rm sca}$, with a difference of more than 0.16 $M_{\odot}$.

In the models of KIC 6131884, $M_0$ and $\eta$ are set to 1.08 $M_{\odot}$ and 0.50, respectively. This value represents the maximum initial mass the star can have, according to the MZ diagram. The models constructed for this star are used to investigate the variations in $\Delta\nu$, $\Delta\Pi_1$, and $T_{\rm eff}$ throughout its evolution. RC2A and RC2B are located near the ZACHeB phase, while RC2C, RC2D, and RC2E correspond to the nTACHeB phase.
Although the $\Delta\nu$ values of RC2A and RC2B lie outside the observational uncertainty limits, RC2A  is close to the value reported by Kallinger (2019). The $\Delta\Pi_1$ uncertainty for KIC 6131884 is not provided in Gaulme et al. (2020). Since the $\Delta\Pi_1$ values of KIC 5526130 and KIC 6131884 are similar, we adopt the same uncertainty of ±9 s. Under this assumption, the $\Delta\Pi_1$ values of the first two models remain higher than the observed value. Also, among the five models, RC2A and RC2B have the highest $\chi^2_{\rm seis}$ values, with 3.95 and 18.31, respectively.

In contrast, the $\Delta\Pi_1$ values in RC2C, RC2D, and RC2E lie between 259.8 and 262.2 s, all of which are in good agreement with the observational value within the adopted uncertainty. Among these, RC2D also shows excellent agreement in $\Delta\nu$ with the observational value. Furthermore, its $T_{\rm eff}$ is 4876 K, which lies within the observational uncertainty range.
Considering $\Delta\nu$, $\Delta\Pi_1$, and $T_{\rm eff}$ simultaneously, RC2D emerges as the best-fitting model for KIC 6131884, with the lowest $\chi^2_{\rm seis}$ value of 0.20 among the six models. $M_{\rm f}$ of this model is very close to the value predicted by $M_{\rm sca}$, with a difference of approximately 0.02 $M_{\odot}$.

Although the best-fitting models have been obtained for RCs, they are not the only solutions. Varying parameters like $M_0$, $\alpha$, and $\eta$ can yield multiple compatible models.

\subsection{Asteroseismic Analysis of Red Giant Stars}
Seismic data show variations depending on the evolutionary phase of RG stars. Table \ref{tab:asteroseismic_obs} contains the observational values of $\Delta\nu$, $\nu_{\rm max}$, and $\Delta\Pi_1$ for the studied stars \citep{2019arXiv190609428K, 2018ApJS..239...32P, 2016A&A...588A..87V, 2020A&A...639A..63G}. The oscillation frequencies of the six stars are provided in \citet{2019arXiv190609428K}. The $\Delta\nu$ values of the models are derived from the $\Delta\nu$–$\nu$ diagram. Therefore, for a consistent comparison between the model and observational values, the observed $\Delta\nu$ values are also determined from the $\Delta\nu$–$\nu$ diagram.  

\begin{figure}
    \centering
    \includegraphics[width=1.0\linewidth]{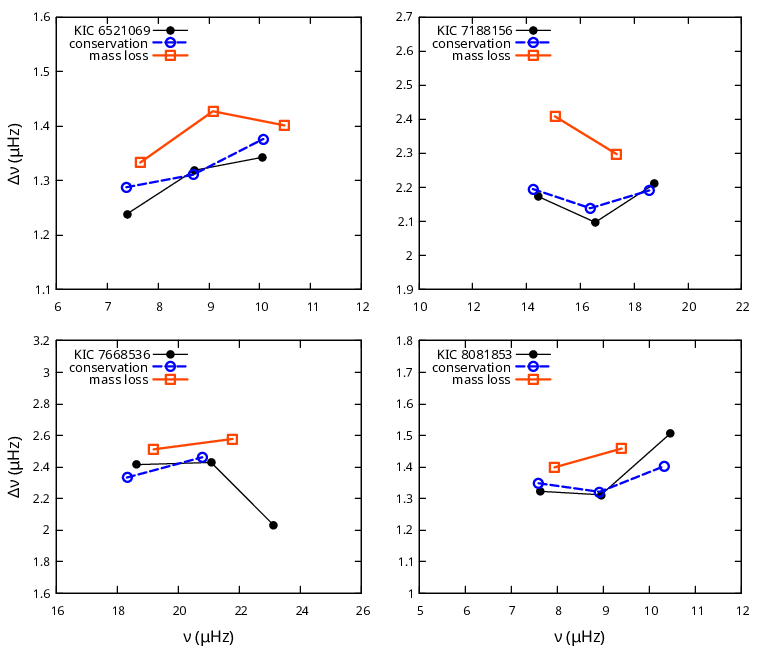}
    \caption{$\Delta\nu-\nu$ plot of $l=0$ frequencies for stars KIC 6521069, KIC 7188156, KIC 7668536 and KIC 8081853. The mass-loss model, mass-conservation model, and observational frequencies are represented by squares, circles, and solid circles, respectively.}
    \label{fig:Dnu-nu-4ML}
\end{figure}
Although the values of $M$, $R$, and $T_{\rm eff}$ for the last four stars in Table \ref{tab:models} are very similar in both the mass-conservation and mass-loss models, their seismic properties differ (see Table \ref{tab:models} and Fig. \ref{fig:Dnu-nu-4ML}). The mass-conservation models show better agreement with the observations on the $\Delta\nu$–$\nu$ diagram than the mass-loss models. However, they cannot be used to represent these stars because their ages are significantly older than the age of the Galaxy. The $\Delta\nu$ values of the mass-loss models are 4–9$\%$ higher than the observed values. The $\Delta\nu$ values of these models exceed the observational uncertainty range. To achieve consistency in $\Delta\nu$, the mean density must be reduced.

The $l=0$ frequencies of the mass-conservation and mass-loss models in the $\Delta\nu$–$\nu$ diagram shown in Fig. \ref{fig:MZ} correspond to the same radial order ($n$). In the mass-loss models of KIC 6521069, KIC 7188156, KIC 7668536, and KIC 8081853, the $l=0$ oscillation frequencies are slightly higher than the observed values. By reducing the mean density, both the frequencies and the $\Delta\nu$ values can be brought into better agreement with the observational data. There are three ways to achieve this: keep $R$ constant and decrease $M$, keep $M$ constant and increase $R$, or adjust both parameters simultaneously.

The observational $l=1$ oscillation frequencies of these four stars could not be determined by \citet{2019arXiv190609428K}. Therefore, the $l=1$ mixed-mode oscillation frequencies of the mass-loss and mass-conservation models are compared. Although the $M$, $R$, and $T_{\rm eff}$ values of the models are the same, differences are found in their $l=1$ frequencies.

\begin{figure}
    \centering
    \includegraphics[width=1.15\linewidth]{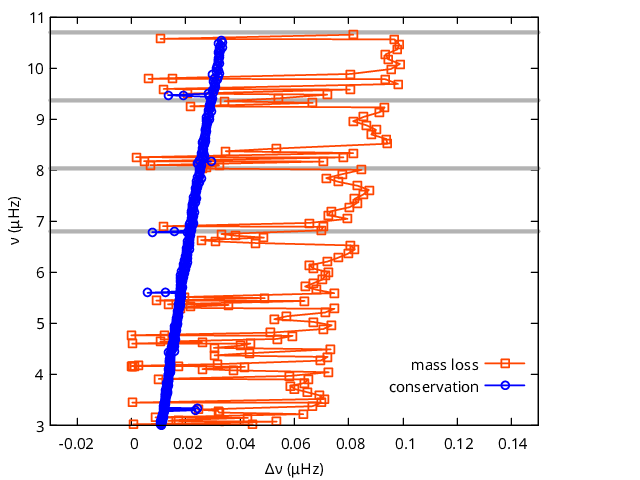}
    \caption{Plot of $l=1$ frequencies $\nu$ against $\Delta\nu$ for KIC 6521069. Squares represent the mass loss model, and circles represent the mass conservation model. The gray solid lines represent the l=1 frequencies calculated from the relationship $\nu_{l=1}=\nu_{l=0}+(\Delta\nu/2)$.}
    \label{fig:l1model}
\end{figure}

Fig. \ref{fig:l1model} shows the $\nu$–$\Delta\nu$ diagram for the $l=1$ frequencies of KIC 6521069. In the mass-conservation model, the successive differences between adjacent $l=1$ frequencies exhibit a regular pattern, with alternating minima and maxima. In contrast, the mass-loss model shows a more scattered structure. Nevertheless, similar minima also appear in the mass-loss model. These minima correspond to the $l=1$ frequencies estimated from the relation $\nu_{l=1} = \nu_{l=0} + (\Delta\nu / 2)$. If the $l=1$ modes are not affected by mode mixing, they would follow this relation. For this reason, we refer to these frequencies as ‘pseudo $l=1$ frequencies’.

\begin{figure}
    \centering
    \includegraphics[width=1.0\linewidth]{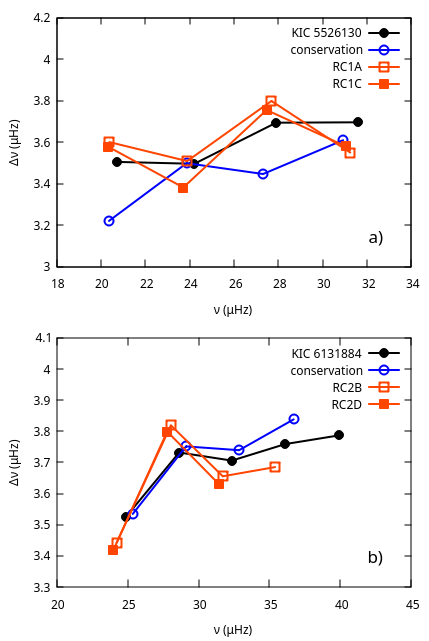}
    \caption{$\Delta\nu$–$\nu$ diagram of $l=0$ frequencies.
a) KIC 5526130: mass-conservation model (circles), RC1A model (open squares), and RC1C model (filled squares) are shown. b) KIC 6131884: mass-conservation model (circles), RC2B model (open squares), and RC2D model (filled squares) are shown. Observed frequencies are represented by filled circles.}
    \label{fig:RC-Dnu-nu}
\end{figure}
The $l=0$ mode frequencies of the RC models are compared with observations in Fig. \ref{fig:RC-Dnu-nu}. In Fig. \ref{fig:RC-Dnu-nu}a, the $\Delta\nu$–$\nu$ pattern of the RC1A model for KIC 5526130 shows better agreement with the observations than the RC1C model. Additionally, its frequencies are closer to the observed values. However, RC1C provides a better overall match when $\Delta\nu$, $\Delta\Pi_1$, and $T_{\rm eff}$ are considered together.
For KIC 6131884, the RC2B and RC2D models are compared with observations in Fig. \ref{fig:RC-Dnu-nu}b. These two models have similar structures. Although the mass-conservation model agrees more closely with the observed frequencies, it cannot represent the star due to its age (14.15 Gyr), which exceeds the age of our galaxy.

\subsection{Individual Notes on Red Giant Stars} 
\subsubsection{KIC 5526130}
KIC 5526130 is classified among the RGB stars in the APOKASC-2 catalogue. The star’s $M_{\rm sca}$, $R_{\rm sca}$, and $Z_{\rm s}$ are 0.6298 $\rm M_{\odot}$, 9.5007 $\rm R_{\odot}$, and 0.0019, respectively. It is located in the region of the HR diagram where RGB and RC stars overlap. Assuming it is on its first ascent, the model age of the star is 24.4 Gyr—significantly older than the age of our galaxy. With this combination of $M$ and $Z$, it is not possible for KIC 5526130 to be on the first RGB ascent.

Consistent with this, KIC 5526130 is located outside the left edge of the triangle in the MZ diagram, indicating clear evidence of mass loss. According to the diagram, the minimum initial mass is 0.93 $\rm M_{\odot}$, and the minimum amount of mass lost is 0.2422 $\rm M_{\odot}$. Since mass loss is primarily effective near the RGB tip, the star must have ascended the RGB and subsequently descended into the RC region. The estimated mass-loss parameter, calculated from the grid models, is $\eta_{\rm min} = 0.4083$ (see Section \ref{sec:M0Zs}).

The $\Delta\Pi_1$ value of this star is reported in the literature as $245.4 \pm 8.94$ s by \citet{2016A&A...588A..87V}. The $\Delta\nu$ value of KIC 5526130 is given as $3.492 \pm 0.019$ $\mu$Hz in \citet{2019arXiv190609428K}, and as $3.567 \pm 0.011$ $\mu$Hz in the APOKASC-2 catalog. According to the criteria proposed by \citet{Mosser2014}, if $\Delta\nu \leq 10$ $\mu$Hz and $\Delta\Pi_1 > 100$ s, the star can be classified as an RC star. Based on these asteroseismic parameters, KIC 5526130 is identified as an RC star. Accordingly, Section \ref{sec:RCmodel} presents the RC models constructed for this star.

\subsubsection{KIC 6131884}
The estimated $M_{\rm sca}$ and $R_{\rm sca}$ values for KIC 6131884, classified as an RGB star in the APOKASC-2 catalogue, are 0.8219 $\rm M_{\odot}$ and 10.0832 $\rm R_{\odot}$, respectively, based on asteroseismic data. The metallicity $Z$, calculated from [Fe/H], is 0.0033. KIC 6131884 is located near the luminosity bump on the $\log(L)$–$\log(T_{\rm eff})$ diagram (Fig.~\ref{fig:LTeff}). Assuming the star is on its first ascent of the RGB, a mass-conservation model can be constructed. However, the age of the model obtained with 0.8219 $\rm M_{\odot}$ is approximately 14.15 Gyr, which exceeds the estimated age of the Galaxy. 

Given the values of $M_{\rm sca}$ and $Z_{\rm s}$, KIC 6131884 lies outside the triangle in the MZ diagram. For the given $Z_{\rm s}$, the minimum initial mass ($M_{0\rm min}$) is 0.93 $\rm M_{\odot}$, corresponding to a minimum mass loss of 0.1081 $\rm M_{\odot}$. The corresponding $\eta_{\rm min}$ value, derived from the grid models, is 0.2077. The base of the triangle represents the maximum possible initial mass the star could have had. The mass at the point where $Z_{\rm s}$ intersects the base of the triangle is $M_{0\rm max} = 1.08$ $\rm M_{\odot}$. In this case, the maximum possible mass lost by the star is $\Delta M_{\rm max} = 0.26$ $\rm M_{\odot}$, and the estimated value of $\eta_{\rm max}$ required to reach this is 0.5696. For both initial mass limits, the star must have climbed the RGB and subsequently descended into the RC region.

The $\Delta\Pi_1$ value of KIC 6131884 is reported as 255.40 s in \citet{2020A&A...639A..63G}. As shown in Table \ref{tab:models}, the observational $\Delta\nu$ values are 3.612 and 3.703 $\mu$Hz. Based on these $\Delta\Pi_1$ and $\Delta\nu$ values, KIC 6131884 is classified as an RC star. Accordingly, RC models have been constructed for this star (see Section \ref{sec:RCmodel}).

\subsubsection{KIC 6521069}
KIC 6521069 is in the RGB phase. Its mass and radius, derived from asteroseismic data, are 0.8331 $\rm M_{\odot}$ and 20.2999 $\rm R_{\odot}$, respectively. The star's effective temperature and metallicity are 4330.9 K and 0.0068. When a mass-conservative model is constructed during the first-ascent RGB phase using these fundamental parameters, the resulting stellar age is approximately 16 Gyr. This value exceeds the estimated age of the Galaxy, $13.4 \pm 0.8$ Gyr. Therefore, the star must have experienced mass loss.

On the MZ diagram, KIC 6521069 lies outside the left edge of the triangle. According to this diagram, the star must have lost at least $\sim$0.14 $\rm M_{\odot}$ of mass. In this case, the minimum initial mass is estimated to be 0.97 $\rm M_{\odot}$.

The radius of KIC 6521069 is approximately 20 $\rm R_{\odot}$. For the star to lose mass effectively at this radius, the mass-loss parameter $\eta$ would need to be as high as $\sim$3. Such a value is unrealistically high for a single star to achieve before reaching the RGB tip. Therefore, KIC 6521069 is expected to have descended into the RC region following the helium flash at the RGB tip. The required $\eta$ value for this evolution is 0.249. The age of the corresponding mass-loss model is 9.5 Gyr, which is more consistent with the estimated age of the Galaxy.

The $\Delta\nu$ value determined from the model’s $\Delta\nu$–$\nu$ diagram is 1.39 $\mu$Hz (Fig. \ref{fig:Dnu-nu-4ML}), while the observational value obtained from the same diagram is 1.30 $\mu$Hz. The model's $\Delta\nu$ is approximately 7$\%$ higher than the observed value. Based on the relationship between $\Delta\nu$ and the mean stellar density, the current mass of the star is estimated to be around 0.7 $\rm M_{\odot}$.  

\subsubsection{KIC 7188156}
On the HR diagram, the star is located just above the luminosity bump. Assuming it is in its first ascent, mass loss through cold stellar winds is only beginning to take effect. Therefore, constructing a mass-conservative model for the star is justifiable. The age of the mass-conservative model, constructed with $M_{\rm sca} = 0.7874$ $\rm M_{\odot}$ and $Z_{\rm s} = 0.0026$, is 15.75 Gyr. At this age, KIC 7188156 cannot be on the first ascent.

Another indicator of this situation, apart from age, is the position of the star on the MZ diagram. For this star, which lies outside the left edge of the triangle, the minimum and maximum mass losses are estimated to be 0.095 $\rm M_{\odot}$ and 0.19 $\rm M_{\odot}$, respectively. The corresponding initial masses are $M_{0\rm min} = 0.92~{\rm M}_{\odot}$ and $M_{0\rm max} = 1.02~{\rm M}_{\odot}$. The $\eta$ and age of the model with $0.92~{\rm M}{\odot}$ are 0.182 and 9.20 Gyr, respectively, while for the $1.02~{\rm M}_{\odot}$ model, these values are 0.416 and 6.44 Gyr. In this way, the lower and upper limits for $M_0$, $\eta$, and age are determined, constraining the range in which the best-fitting model is expected to lie. Although the initial masses differ, the models yield the same $\Delta\nu$ and $\nu_{\rm max}$ values because their final $M$, $R$, and $T_{\rm eff}$ values are identical.

\subsubsection{KIC 7668536}
KIC 7668536, listed as an RGB star in the APOKASC-2 catalogue, has a mass and radius of 0.7915 $\rm M_{\odot}$ and 13.4027 $\rm R_{\odot}$, respectively. Its Z is 0.0098. On the $\log(L)$–$\log(T_{\rm eff})$ diagram (Fig. \ref{fig:LTeff}), the star is located just above the RC region. A mass-conservative model constructed with 0.7915 $\rm M_{\odot}$ yields an age exceeding 20 Gyr. Given the age of the Galaxy, KIC 7668536 cannot reach the position shown in Fig. \ref{fig:LTeff} with this mass and metallicity. The star must have undergone mass loss.

In Fig. \ref{fig:MZ}, the star lies outside the triangle based on its $M$ and $Z$, which is another indication that it has undergone mass loss. The minimum initial mass determined from the diagram is 1.02 $\rm M_{\odot}$, with a corresponding $\eta$ value of 0.419. The age of the mass-loss model constructed using these parameters is 9.097 Gyr. The model’s $\Delta\nu$ value is approximately 5$\%$ higher than the observational value determined from the $\Delta\nu$–$\nu$ diagram. 

{Another possibility is that this star is in the asymptotic giant branch (AGB) phase. \cite{2019MNRAS.489.4641E} investigated the evolutionary stages of evolved stars using different methods and concluded that KIC 7668536 could be either RGB or AGB. \cite{2019arXiv190609428K} and \cite{2021A&A...650A.115D}, on the other hand, identified KIC 7668536 as an AGB star.}

\subsubsection{KIC 8081853}
KIC 8081853 is classified as an RGB star in the APOKASC-2 catalog. The mass and radius of the star, calculated using asteroseismic parameters, are $M_{\rm sca} = 0.7425~\rm M_{\odot}$ and $R_{\rm sca} = 19.1340~\rm R_{\odot}$, respectively. A mass-conservative model constructed under the assumption that the star is on its first ascent yields an age of 26.825 Gyr. Although the model’s $\Delta\nu$ and oscillation frequencies are consistent with observations, the age clearly indicates that this model cannot represent the star.

In the MZ diagram, the minimum initial mass the star can have is determined to be 1.01~$\rm M_{\odot}$. The required mass loss is approximately 0.27~$\rm M_{\odot}$, corresponding to an $\eta$ value of 0.4695. The age of the model constructed with these parameters is 9.3 Gyr. The model's $\Delta\nu$ is 8$\%$ higher than the observational $\Delta\nu$ determined from the $\Delta\nu$–$\nu$ diagram, indicating that the mean density must be reduced. 


\section{Conclusions}
Red giants enrich the interstellar medium by losing mass during their ascent on the RGB. However, both the amount of mass they lose during this phase and their initial masses remain unresolved issues.
First, we investigate which parameters other than $\eta$ influence mass loss. For this purpose, grids are constructed with $M_0$ in the range 1.0–1.4, $Z$ in the range 0.005–0.035, and $\eta$ values of 0.20, 0.50, and 0.85.
Examinations of the grids reveal that mass loss depends on both initial mass and $Z$. An increase in $Z$ leads to greater mass loss, while an increase in $M_0$ results in reduced mass loss. In this case, stars may exhibit different $\eta$ values depending on their $M$, $Z$, and $\Delta M$.

In this study, we examine six low-mass, low-metallicity stars classified as RGB stars in the APOKASC-2 Catalogue. When these stars are assumed to be in their first ascent and corresponding mass-conservation models are constructed, the resulting model ages are found to be significantly older than the estimated age of the Galaxy (13.4 Gyr). Therefore, these stars cannot be in the first-ascent phase. They must have reached the tip of the RGB, undergone mass loss, and begun their descent toward the RC phase.

Another factor supporting mass loss is that these stars lie outside the left edge of the triangle in the MZ diagram. This edge represents the minimum mass a star must have to reach the RGB region. Therefore, these six stars must have undergone mass loss. The minimum mass loss estimated from the MZ diagram for these stars ranges from 0.1 to 0.3 ${\rm M}_{\odot}$, while the corresponding minimum $\eta$ values lie between 0.20 and 0.50.
Among these stars, KIC 5526130 and KIC 6131884 are located in the region where the RC and RGB phases overlap in the HR diagram. According to their $\Delta\Pi_1$ and $\Delta\nu$ values, both stars have entered the RC phase. The remaining four stars lie above the RC region in the HR diagram.

The ages of the mass-loss models are more consistent with the age of the Galaxy, falling within the range of 9–9.5 Gyr. The ages of the models with the minimum initial mass represent the maximum possible age the stars can have. Conversely, the $M_0$ value determined from the base of the triangle corresponds to the minimum possible stellar age. The maximum initial mass and $\eta$ values have been determined for the RC star KIC 6131884 and the RGB-descending star KIC 7188156. Using these parameters, the ages of the resulting models are 5.67 and 6.44 Gyr, respectively. These values represent the minimum possible ages of the stars.

The degeneracy in the HR diagram makes it difficult to construct interior models of RC stars. In this context, asteroseismic parameters are particularly useful for modeling RC stars. It is possible to distinguish models using $\Delta\nu$, $\Delta\Pi_1$, and $T_{\rm eff}$. A coherent model has been obtained for KIC 5526130 and KIC 6131884 using these three parameters. However, it is difficult to claim that these models are unique, as multiple coherent models can be produced by varying the input parameters. Additional constraining parameters are needed to construct a unique model.

The individual frequencies and $\Delta\nu$ values of the mass-loss models are higher than the observational values. In contrast, the seismic data of the mass-conservation models with similar $M$, $R$, and $T_{\rm eff}$ show better agreement with observations. This discrepancy may arise from differences in the internal structures of stars on the RGB descent compared to those on the first ascent. By reducing the mean density, the agreement between the $\Delta\nu$ values and individual frequencies of the mass-loss models and the observations can be improved.

While $l=1$ frequencies are detected in stars located below the upper boundary of the RC region, they are not detected in stars above this limit. Therefore, we compare the $l=1$ frequencies of the models. During the first ascent, $\Delta\nu$ varies within a narrower range, whereas during the descent, it spans a wider range. In both models, minimum values appear at regular intervals. These minima correspond to the pseudo $l=1$ frequencies. If the observational $l=1$ frequencies can be detected in stars within this region, it may be possible to distinguish between ascending and descending RGB stars.

\section*{Acknowledgements}
This work is supported by the Scientific and Technological Research Council of Turkey (TÜBİTAK: 123F019).

\section*{Data Availability}

 The data underlying this article will be shared on reasonable request to the corresponding author.




\begin{thebibliography}{99}
\bibitem[\protect\citeauthoryear{Abdurro'uf et al.}{2022}]{2022ApJS..259...35A} Abdurro'uf, Accetta K., Aerts C., Silva Aguirre V., Ahumada R., Ajgaonkar N., Filiz Ak N., et al., 2022, ApJS, 259, 35. doi:10.3847/1538-4365/ac4414
\bibitem[\protect\citeauthoryear{Angulo et al.}{1999}]{Angulo1999}
Angulo, C., Arnould, M., Rayet, M., Descouvemont, P., Baye, D., et al. 1999, NuPhA, 656, 3
\bibitem[\protect\citeauthoryear{Asplund et al.}{2009}]{2009ARA&A..47..481A} Asplund M., Grevesse N., Sauval A.~J., Scott P., 2009, ARA\&A, 47, 481. doi:10.1146/annurev.astro.46.060407.145222
\bibitem[\protect\citeauthoryear{Baglin et al.}{2006}]{Baglin2006}
Baglin A., Michel E., Auvergne M., COROT Team, 2006, in Fridlund M., Baglin A., Lochard J., Conroy L., eds, ESA Special Publication Vol. 1306, ESA Special Publication. p. 33
\bibitem[\protect\citeauthoryear{Bedding et al.}{2011}]{Bedding2011}
Bedding, T.R., Mosser, B., Huber, D., Montalb$\rm\acute{a}$n, J., Beck, P., et al. 2011, Nature, 471,608
\bibitem[\protect\citeauthoryear{Belkacem et al.}{2011}]{2011A&A...530A.142B} Belkacem K., Goupil M.~J., Dupret M.~A., Samadi R., Baudin F., Noels A., Mosser B., 2011, A\&A, 530, A142. doi:10.1051/0004-6361/201116490
\bibitem[\protect\citeauthoryear{Borucki et al.}{2010}]{Borucki2010}
Borucki W. J., Koch D., Basri G., Batalha, N., Broen, T., et al. 2010, Science, 327, 977
\bibitem[\protect\citeauthoryear{Brogaard et al.}{2024}]{2024arXiv241008330B} Brogaard K., Miglio A., van Rossem W.~E., Willett E., Thomsen J.~S., 2024, arXiv, arXiv:2410.08330. doi:10.48550/arXiv.2410.08330
\bibitem[\protect\citeauthoryear{Brown}{1991}]{Brown1991}
Brown, T.M., Gilliland R.L., Noyes, R.W., Ramsey, L.W., 1991, ApJ, 368, 599
\bibitem[\protect\citeauthoryear{Christensen-Dalsgaard}{1988}]{Christensen1988}
Christensen-Dalsgaard J., 1988, in Advances in Helio- and Asteroseismology , Aarhus, Denmark, July 11-17, 1986. Editors Christensen-Dalsgaard J. and Frandsen, S., IAU Symp. 123, Dordrecht: D. Reidel Publishing Co., p.295
\bibitem[\protect\citeauthoryear{Christensen-Dalsgaard}{2008}]{Christensen2008}
Christensen-Dalsgaard J., 2008, ApSS, 316, 113
{\bibitem[\protect\citeauthoryear{Christensen-Dalsgaard et al.}{2010}]{2010ApJ...713L.164C} Christensen-Dalsgaard J., Kjeldsen H., Brown T.~M., Gilliland R.~L., Arentoft T., Frandsen S., Quirion P.-O., et al., 2010, ApJL, 713, L164. doi:10.1088/2041-8205/713/2/L164}
\bibitem[\protect\citeauthoryear{Christensen-Dalsgaard}{2015}]{2015MNRAS.453..666C} Christensen-Dalsgaard J., 2015, MNRAS, 453, 666. doi:10.1093/mnras/stv1656
{\bibitem[\protect\citeauthoryear{Creevey et al.}{2007}]{2007ApJ...659..616C} Creevey O.~L., Monteiro M.~J.~P.~F.~G., Metcalfe T.~S., Brown T.~M., Jim{\'e}nez-Reyes S.~J., Belmonte J.~A., 2007, ApJ, 659, 616. doi:10.1086/512097}
{\bibitem[\protect\citeauthoryear{Dr{\'e}au et al.}{2021}]{2021A&A...650A.115D} Dr{\'e}au G., Mosser B., Lebreton Y., Gehan C., Kallinger T., 2021, A\&A, 650, A115. doi:10.1051/0004-6361/202040240}
{\bibitem[\protect\citeauthoryear{Elsworth et al.}{2019}]{2019MNRAS.489.4641E} Elsworth Y., Hekker S., Johnson J.~A., Kallinger T., Mosser B., Pinsonneault M., Hon M., et al., 2019, MNRAS, 489, 4641. doi:10.1093/mnras/stz2356}
{\bibitem[\protect\citeauthoryear{Ferguson et al.}{2005}]{Ferguson2005} Ferguson, J.W., Alexander, D.R., Allard, F., Barmanu, T., et al. 2005, ApJ, 623, 585}
\bibitem[\protect\citeauthoryear{Gaulme et al.}{2020}]{2020A&A...639A..63G} Gaulme P., Jackiewicz J., Spada F., Chojnowski D., Mosser B., McKeever J., Hedlund A., et al., 2020, A\&A, 639, A63. doi:10.1051/0004-6361/202037781
\bibitem[\protect\citeauthoryear{Gratton et al.}{2003}]{2003A&A...408..529G} Gratton R.~G., Bragaglia A., Carretta E., Clementini G., Desidera S., Grundahl F., Lucatello S., 2003, A\&A, 408, 529. doi:10.1051/0004-6361:20031003
\bibitem[\protect\citeauthoryear{Handberg et al.}{2017}]{2017MNRAS.472..979H} Handberg R., Brogaard K., Miglio A., Bossini D., Elsworth Y., Slumstrup D., Davies G.~R., et al., 2017, MNRAS, 472, 979. doi:10.1093/mnras/stx1929
\bibitem[\protect\citeauthoryear{Hekker et al.}{2020}]{2020MNRAS.492.5940H} Hekker S., Angelou G.~C., Elsworth Y., Basu S., 2020, MNRAS, 492, 5940. doi:10.1093/mnras/staa176
\bibitem[\protect\citeauthoryear{Howell et al.}{2022}]{2022MNRAS.515.3184H} Howell M., Campbell S.~W., Stello D., De Silva G.~M., 2022, MNRAS, 515, 3184. doi:10.1093/mnras/stac1918
\bibitem[\protect\citeauthoryear{Huber et al.}{2011}]{Huber2011} Huber D., Bedding T.R., Stello D., et al., 2011, ApJ, 743, 143
\bibitem[\protect\citeauthoryear{Huber et al.}{2017}]{2017ApJ...844..102H} Huber D., Zinn J., Bojsen-Hansen M., Pinsonneault M., Sahlholdt C., Serenelli A., Silva Aguirre V., et al., 2017, ApJ, 844, 102. doi:10.3847/1538-4357/aa75ca
{\bibitem[\protect\citeauthoryear{Iglesias \& Rogers}{1993}]{Iglesias1993} Iglesias C.~A., Rogers F.~J., 1993, ApJ, 412, 752. doi:10.1086/172958
\bibitem[\protect\citeauthoryear{Iglesias \& Rogers}{1996}]{Iglesias1996} Iglesias C.A., and Rogers F.J. 1996, ApJ, 464, 943}
\bibitem[\protect\citeauthoryear{Jermyn et al.}{2023}]{Jermyn2023} Jermyn A.~S., Bauer E.~B., Schwab J., Farmer R., Ball W.~H., Bellinger E.~P., Dotter A., et al., 2023, ApJS, 265, 15. doi:10.3847/1538-4365/acae8d
{\bibitem[\protect\citeauthoryear{Kallinger et al.}{2010}]{2010A&A...522A...1K} Kallinger T., Mosser B., Hekker S., Huber D., Stello D., Mathur S., Basu S., et al., 2010, A\&A, 522, A1. doi:10.1051/0004-6361/201015263}
{\bibitem[\protect\citeauthoryear{Kallinger et al.}{2012}]{2012A&A...541A..51K} Kallinger T., Hekker S., Mosser B., De Ridder J., Bedding T.~R., Elsworth Y.~P., Gruberbauer M., et al., 2012, A\&A, 541, A51. doi:10.1051/0004-6361/201218854}
\bibitem[\protect\citeauthoryear{Kallinger}{2019}]{2019arXiv190609428K} Kallinger T., 2019, arXiv, arXiv:1906.09428. doi:10.48550/arXiv.1906.09428
\bibitem[\protect\citeauthoryear{Kjeldsen \& Bedding }{1995}]{Kjebed1995} 
Kjeldsen H., and Bedding T. R. 1995, A\&A, 293, 87
\bibitem[\protect\citeauthoryear{Kjeldsen, Bedding \& Christensen-Dalsgard}{2008}]{KBC2008} 
Kjeldsen H., Bedding T. R., Christensen-Dalsgard J., 2008, ApJL, 683, L175
\bibitem[\protect\citeauthoryear{Lebzelter \& Wood}{2011}]{2011A&A...529A.137L} Lebzelter T., Wood P.~R., 2011, A\&A, 529, A137. doi:10.1051/0004-6361/201016319
\bibitem[\protect\citeauthoryear{Li et al.}{2022}]{2022NatAs...6..673L} Li Y., Bedding T.~R., Murphy S.~J., Stello D., Chen Y., Huber D., Joyce M., et al., 2022, NatAs, 6, 673. doi:10.1038/s41550-022-01648-5
\bibitem[\protect\citeauthoryear{Li et al.}{2022}]{2022ApJ...927..167L} Li T., Li Y., Bi S., Bedding T.~R., Davies G., Du M., 2022, ApJ, 927, 167. doi:10.3847/1538-4357/ac4fbf
\bibitem[\protect\citeauthoryear{McDonald \& Zijlstra}{2015}]{2015MNRAS.448..502M} McDonald I., Zijlstra A.~A., 2015, MNRAS, 448, 502. doi:10.1093/mnras/stv007
\bibitem[\protect\citeauthoryear{Miglio et al.}{2012}]{2012MNRAS.419.2077M} Miglio A., Brogaard K., Stello D., Chaplin W.~J., D'Antona F., Montalb{\'a}n J., Basu S., et al., 2012, MNRAS, 419, 2077. doi:10.1111/j.1365-2966.2011.19859.x
\bibitem[\protect\citeauthoryear{Miglio et al.}{2021}]{2021A&A...645A..85M} Miglio A., Chiappini C., Mackereth J.~T., Davies G.~R., Brogaard K., Casagrande L., Chaplin W.~J., et al., 2021, A\&A, 645, A85. doi:10.1051/0004-6361/202038307
\bibitem[\protect\citeauthoryear{Mosser et al.}{2014}]{Mosser2014} 
Mosser, B., Benomar, O., Belkacem, K., et al. 2014, A\&A, 572, L5
\bibitem[\protect\citeauthoryear{Paquette et al.}{1986}]{1986ApJS...61..177P} Paquette C., Pelletier C., Fontaine G., Michaud G., 1986, ApJS, 61, 177. doi:10.1086/191111
\bibitem[\protect\citeauthoryear{Paxton et al.}{2011}]{Paxton2011} 
Paxton B., Bildsten L., Dotter A., Herwig F., Lesaffre P. and Timmes F., 2011, ApJS, 192, 35
\bibitem[\protect\citeauthoryear{Paxton et al.}{2013}]{Paxton2013} 
Paxton B., Cantiello M., Arras P., Bildsten L., Brown, E.F, et al., 2013, ApJS, 208, 42
\bibitem[\protect\citeauthoryear{Paxton et al.}{2015}]{Paxton2015} Paxton B., Marchant P., Schwab J., Bauer E.~B., Bildsten L., Cantiello M., Dessart L., et al., 2015, ApJS, 220, 15. doi:10.1088/0067-0049/220/1/15
\bibitem[\protect\citeauthoryear{Paxton et al.}{2018}]{Paxton2018} Paxton B., Schwab J., Bauer E.~B., Bildsten L., Blinnikov S., Duffell P., Farmer R., et al., 2018, ApJS, 234, 34. doi:10.3847/1538-4365/aaa5a8
\bibitem[\protect\citeauthoryear{Paxton et al.}{2019}]{Paxton2019} Paxton B., Smolec R., Schwab J., Gautschy A., Bildsten L., Cantiello M., Dotter A., et al., 2019, ApJS, 243, 10. doi:10.3847/1538-4365/ab2241
\bibitem[\protect\citeauthoryear{Pinsonneault et al.}{2018}]{2018ApJS..239...32P} Pinsonneault M.~H., Elsworth Y.~P., Tayar J., Serenelli A., Stello D., Zinn J., Mathur S., et al., 2018, ApJS, 239, 32. doi:10.3847/1538-4365/aaebfd
\bibitem[\protect\citeauthoryear{Planck Collaboration et al.}{2020}]{2020A&A...641A...6P} Planck Collaboration, Aghanim N., Akrami Y., Ashdown M., Aumont J., Baccigalupi C., Ballardini M., et al., 2020, A\&A, 641, A6. doi:10.1051/0004-6361/201833910
\bibitem[\protect\citeauthoryear{Pr{\v{s}}a et al.}{2016}]{Prsa2016} Pr{\v{s}}a A., Harmanec P., Torres G., Mamajek E., Asplund M., Capitaine N., Christensen-Dalsgaard J., et al., 2016, AJ, 152, 41. doi:10.3847/0004-6256/152/2/41
\bibitem[\protect\citeauthoryear{Rauer et al.}{2014}]{2014ExA....38..249R} Rauer H., Catala C., Aerts C., Appourchaux T., Benz W., Brandeker A., Christensen-Dalsgaard J., et al., 2014, ExA, 38, 249. doi:10.1007/s10686-014-9383-4
\bibitem[\protect\citeauthoryear{Reimers}{1975}]{1975MSRSL...8..369R} Reimers D., 1975, MSRSL, 8, 369
\bibitem[\protect\citeauthoryear{Salaris, Cassisi, \& Pietrinferni}{2016}]{2016A&A...590A..64S} Salaris M., Cassisi S., Pietrinferni A., 2016, A\&A, 590, A64. doi:10.1051/0004-6361/201628181
\bibitem[\protect\citeauthoryear{Sharma et al.}{2016}]{2016ApJ...822...15S} {Sharma S., Stello D., Bland-Hawthorn J., Huber D., Bedding T.~R., 2016, ApJ, 822, 15. doi:10.3847/0004-637X/822/1/15}
{\bibitem[\protect\citeauthoryear{Silva Aguirre et al.}{2015}]{2015MNRAS.452.2127S} Silva Aguirre V., Davies G.~R., Basu S., Christensen-Dalsgaard J., Creevey O., Metcalfe T.~S., Bedding T.~R., et al., 2015, MNRAS, 452, 2127. doi:10.1093/mnras/stv1388}
\bibitem[\protect\citeauthoryear{Sullivan et al.}{2015}]{Sullivan2015} 
Sullivan, P. W., Winn, J.N., Berta-Thompson, Z.K., Charbonneau, D., Deming, D., et al., 2015, ApJ, 809, 77
\bibitem[\protect\citeauthoryear{Tailo et al.}{2020}]{2020MNRAS.498.5745T} Tailo M., Milone A.~P., Lagioia E.~P., D'Antona F., Marino A.~F., Vesperini E., Caloi V., et al., 2020, MNRAS, 498, 5745. doi:10.1093/mnras/staa2639
\bibitem[\protect\citeauthoryear{Ulrich}{1986}]{Ulrich1986} 
Ulrich, R.K., 1986, ApJ, 306, L37
\bibitem[\protect\citeauthoryear{Valle et al.}{2018}]{2018A&A...609A..58V} Valle G., Dell'Omodarme M., Prada Moroni P.~G., Degl'Innocenti S., 2018, A\&A, 609, A58. doi:10.1051/0004-6361/201730880
\bibitem[\protect\citeauthoryear{Viani et al.}{2017}]{2017ApJ...843...11V} Viani L.~S., Basu S., Chaplin W.~J., Davies G.~R., Elsworth Y., 2017, ApJ, 843, 11. doi:10.3847/1538-4357/aa729c
\bibitem[\protect\citeauthoryear{Vrard, Mosser, \& Samadi}{2016}]{2016A&A...588A..87V} Vrard M., Mosser B., Samadi R., 2016, A\&A, 588, A87. doi:10.1051/0004-6361/201527259
\bibitem[\protect\citeauthoryear{Wenger et al.}{2000}]{2000A&AS..143....9W} Wenger M., Ochsenbein F., Egret D., Dubois P., Bonnarel F., Borde S., Genova F., et al., 2000, A\&AS, 143, 9. doi:10.1051/aas:2000332
\bibitem[\protect\citeauthoryear{White et al.}{2011}]{2011ApJ...743..161W} White T.~R., Bedding T.~R., Stello D., Christensen-Dalsgaard J., Huber D., Kjeldsen H., 2011, ApJ, 743, 161. doi:10.1088/0004-637X/743/2/161
\bibitem[\protect\citeauthoryear{Y{\i}ld{\i}z et al.}{2016}]{2016MNRAS.462.1577Y} {Y{\i}ld{\i}z M., {\c{C}}elik Orhan Z., Kayhan C., 2016, MNRAS, 462, 1577. doi:10.1093/mnras/stw1709}
\bibitem[\protect\citeauthoryear{Y{\i}ld{\i}z \& {\"O}rtel}{2021}]{YildizOrtel2021} Y{\i}ld{\i}z M., {\"O}rtel S., 2021, MNRAS, 504, 2273. doi:10.1093/mnras/stab996
\bibitem[\protect\citeauthoryear{Y{\i}ld{\i}z}{2023}]{Yıldız2023} Y{\i}ld{\i}z M., 2023, MNRAS, 518, 5552. doi:10.1093/mnras/stac3464
\bibitem[\protect\citeauthoryear{Yu et al.}{2021}]{2021MNRAS.501.5135Y} Yu J., Hekker S., Bedding T.~R., Stello D., Huber D., Gizon L., Khanna S., et al., 2021, MNRAS, 501, 5135. doi:10.1093/mnras/staa3970

\end{thebibliography}



\newpage ~~~ \\
\newpage
\appendix

\section{Inlist for Mass Loss Model}
The \texttt{inlist} used in the MESA evolution code (r23.05.01) for the mass loss model of KIC 8081853 (as given in Table \ref{tab:models}) is provided below. It contains the basic input parameters used in our models, such as the initial mass, metallicity, mixing-length parameter, and efficiency parameter.

\begin{lstlisting}
/&star_job

    mesa_dir = '/home/user/mesa'
    history_columns_file = 'history_columns.list'
    create_pre_main_sequence_model = .true.

    save_model_when_terminate = .true.
    save_model_filename = 'solar_calib.mod'
    filename_for_profile_when_terminate = 'final_profile.data'

    change_initial_net = .true.
    new_net_name = 'pp_and_cno_extras.net'
     
    change_initial_Y = .true.
    change_initial_Z = .true.
    new_Y = 0.2663
    new_Z = 0.0096
    initial_zfracs = 6      
    save_pulse_data_for_model_number = 1000
    save_pulse_data_when_terminate = .true.
    save_pulse_data_filename = 'fgong'

/ !end of star_job namelist

/&eos
/ ! end of eos namelist

/&kap

    kap_file_prefix ='a09' 
    kap_lowT_prefix = 'lowT_fa05_a09p'

/ ! end of kap namelist

/&controls

    initial_mass =  1.01   
    initial_z = 0.0096
    mixing_length_alpha =1.9061
    max_age = 9.326d9

    star_history_name = 'history.data'

    history_interval = 10
    profile_interval = 50
    terminal_interval = 1000
    write_header_frequency = 100000
 
    convergence_ignore_equL_residuals = .true.

    num_trace_history_values = 2
    trace_history_value_name(1) = 'rel_E_err'
    trace_history_value_name(2) = 'log_rel_run_E_err'

    am_nu_visc_factor = 0
    am_D_mix_factor = 0.0333333333333333d0
    D_DSI_factor = 0
    D_SH_factor = 1
    D_SSI_factor = 1
    D_ES_factor = 1
    D_GSF_factor = 1
    D_ST_factor = 1

    varcontrol_target = 1d-3
    delta_lgL_He_limit = 0.01d0

    cool_wind_full_on_T = 9.99d9
    hot_wind_full_on_T = 1d10

    cool_wind_RGB_scheme = 'Reimers'
    Reimers_scaling_factor = 0.469  
 
    atm_option = 'table'
    atm_table = 'photosphere'
 
    show_diffusion_info = .true.
    do_element_diffusion = .true.

    write_pulse_data_with_profile = .false.
    pulse_data_format = 'FGONG'
    fgong_ivers = 300   
    format_for_OSC_data = '(1P5E19.12,x)' 

    calculate_Brunt_N2 = .true.
    brunt_N2_coefficient = 1
    num_cells_for_smooth_brunt_B = 2
    interpolate_rho_for_pulse_data = .true.
    use_other_brunt = .false. 
    min_magnitude_brunt_B = -1d99
 
    max_years_for_timestep =0.5d7           
    varcontrol_target = 1d-4
 
    mesh_delta_coeff = 0.8   	
    P_function_weight = 25
    T_function1_weight = 75
/ ! end of controls namelist

\end{lstlisting}
%

\bsp	
\label{lastpage}
\end{document}